\documentclass[journal=acscatal,manuscript=article]{achemso}
\makeatletter
\newlength{\TOCentrywidth}
\newlength{\TOCentryheight}
\makeatother
\setkeys{acs}{maxauthors = 20}
\usepackage{chemformula}
\usepackage{booktabs}
\usepackage{graphicx}
\usepackage{float}
\usepackage{amsmath}
\usepackage{amssymb}
\usepackage{subcaption}
\usepackage{url}
\usepackage{verbatim}

\author{Jingwen Zhou}
\email{jwzhou@smail.nju.edu.cn}
\affiliation{
	State Key Laboratory of Coordination Chemistry, 
	Key Laboratory of Mesoscopic Chemistry of Ministry of Education, 
	School of Chemistry, Nanjing University, Nanjing 210023, China
}

\author{Yawen Yu}
\affiliation{
	State Key Laboratory of Coordination Chemistry, 
	Key Laboratory of Mesoscopic Chemistry of Ministry of Education, 
	School of Chemistry, Nanjing University, Nanjing 210023, China
}

\author{Xuwei Liu}
\affiliation{
	Frontier Research Department, Baidu Inc., Beijing 100085, China
}

\author{Chungen Liu}
\email{cgliu@nju.edu.cn}
\affiliation{
	State Key Laboratory of Coordination Chemistry, 
	Key Laboratory of Mesoscopic Chemistry of Ministry of Education, 
	School of Chemistry, Nanjing University, Nanjing 210023, China
}

\title[An \textsf{achemso} demo]
  {Explicit Electric Potential-Embedded Machine Learning Framework: A Unified Description from Atomic to Electronic Scales}


\begin{document}

\begin{abstract} 

	To further develop accurate and large-scale theoretical simulations of electrochemical interfaces, we propose a unified explicit electric potential framework to simultaneously predict atomic forces and electron density distributions in electrochemical solid-liquid interface simulations. 
	The framework consists of three components: data generation, model training, and model application. The data generation component is controlled by the Hy\_DFT software, designed to efficiently regulate the potential during constant-potential ab initio molecular dynamics (CP-AIMD) simulations, thereby reducing the number of single-point calculations required for potential convergence. 
	The model training component comprises two decoupled modules: Potential-Embedded MACE (PE-MACE) and Potential-Embedded Electron Density Prediction (PE-EDP). PE-MACE is our implementation of the proposed explicit electric potential machine learning force field (EEP-MLFF) based on the MACE graph neural network software.
	In our previous work, we implemented constant-potential machine learning molecular dynamics (CP-MLMD) simulations based on the potential-embedding method. Building upon this foundation, we further developed the PE-EDP model to address the limitation of EEP-MLFF in describing only nuclear motion. 
	PE-EDP is also constructed on equivariant graph neural networks, aimed at training accurate models capable of predicting electron density distributions under arbitrary potential conditions.
	We employ the Pt(111)/water interface as a representative system to validate our models. Both PE-MACE and PE-EDP demonstrate high accuracy on training and test sets. The radial distribution functions (RDFs) from CP-MLMD simulations at various potentials show good agreement with CP-AIMD results, and long-timescale CP-MLMD simulations reveal the potential-induced reorganization of interface water. The planar-integrated charges along the z-direction and Bader charge analysis from PE-EDP are in good agreement with DFT results.
	These findings demonstrate that our explicit electric potential-embedded unified framework can describe both the atomic-scale dynamics and electronic-scale electron density distribution characteristics of electrochemical interfaces, providing a useful tool for investigating electrochemical interfacial reactions.
\end{abstract}

\section{Introduction}%
	Electrified solid–liquid interfaces are central to a wide range of electrochemical processes, especially electrocatalysis and rechargeable batteries where the interactions between electrode surfaces, explicit solvent structure, and electric potential determine the device performance or the overall efficiency and selectivity of electrochemical reactions. \cite{Liu2016a, Stamenkovic2017,Zachman2018, Zhang2024a} However, because experimental characterization under operating conditions demands sophisticated in situ and operando techniques, these complex coupling mechanism remains poorly understood. Consequently, accurate theoretical modeling of electrochemical interfaces under operando conditions has become essential for elucidating interfacial reaction mechanisms and optimizing device performance. \cite{Yang2022, Yao2022}
	
	Recently, grand canonical density functional theory (GC-DFT) based on the implicit solvation model for electrochemical systems has experienced significant development and widespread application. \cite{Bonnet2012, Bouzid2017, Zhao2021, Liu2020a, Xia2023, Zhou2023}
	Despite these advances, conventional implicit solvation approaches such as VASPsol's linear polarizable continuum model (LPCM) rely on purely local cavity definitions that ignore the finite size of water molecules and solvated ions,
	which can lead to unphysical penetration of cavity into regions too small to accommodate an entire water molecule or solvated ion in some case. \cite{Islam2023}
	The phenomenon is commonly refered as "solvent leakage" in hybrid explicit-implicit solvation model.\cite{Mathew2014, Mathew2019, Islam2023}
	The recently developed VASPsol++ package employed a nonlocal definition of cavity functions to effectively mitigate the "solvent leakage", while incorporating nonlinear responses,\cite{Islam2023} facilitating more accurate and realistic investigations of the charged electrochemical interfaces, especially in constant-potential AIMD (CP-AIMD) simulation.
	
	However, DFT's computational cost severely limits large-scale and long-timescale simulations. Machine learning force fields (MLFFs) have been developed to overcome these computational bottlenecks.
	\cite{Scandolo2019,Bartok2010, Behler2015, Unke2021a,Behler2021}
	Several recent studies have proposed advanced MLFFs specifically designed for electrochemical systems.\cite{Chen2023, Zhou2025,Wang2025, Sun2025, Chen2026} 
	However, some of these models require multiple networks or multiple outputs to predict charges or potentials, resulting in relatively complex architectures that increase both training costs and computational overhead during inference.\cite{Chen2023, Sun2025, Chen2026}
	In addition, approaches that treat the total electron number ($N_e$) as a dynamical variable and propagate its time evolution through a fictitious potentiostat\cite{Bonnet2012} or grand-canonical hybrid Monte Carlo (GC-HMC) sampling\cite{Sun2025} can introduce substantial electric potential fluctuations, \cite{Bonnet2012, Wang2025, Sun2025, Chen2026} 
	which is an artifact of the simulation. \cite{Bonnet2012}
	
	In contrast, our recently proposed EEP-MLFF (Explicit Electric Potential Machine Learning Force Field) framework directly incorporates electric potential as an input parameter to predict system energies and atomic forces, avoiding the extra potential optimization procedure. \cite{Zhou2025} 
	The underlying assumption is that the electron reservoir relaxes instantaneously at each ionic time step to maintain the preset electric potential.
	This treatment may neglect a portion of the genuine thermodynamic fluctuations, but it directly predicts atomic forces at a given potential and propagates the structural evolution without non-physical potential oscillations, serving as an algorithmic approximation aimed at improving simulation stability and efficiency.
	Additionally, for static constant-potential calculations such as geometry optimization and transition-state searches, direct constant-potential optimization is furthermore the most efficient and physically sound approach.

	However, despite the effectiveness of EEP-MLFF in facilitating long-timescale and large-scale simulations at the atomic level, EEP-MLFF cannot predict electron density distributions under arbitrary potentials, limiting electronic-level analysis of electrochemical interface properties and reaction mechanisms.
	Although equivariant graph neural network methods have made progress in electron density prediction, these methods do not account for electric potential information, limiting their direct applicability to electrochemical systems.\cite{jorgensen_equivariant_2022-1, Cheng2023infgcn, kim2024gaussian, fu2024recipe}
	Jiang et al.\ developed an electron density prediction method based on the finite-field model,\cite{Feng2025, Dufils2019} which differs fundamentally from the grand-canonical paradigm adopted here, where the electric potential is controlled by adjusting the electron number of the system to allow electron exchange with an external reservoir.
	However, machine learning models that can predict the real-space electron density distribution of electrochemical interfaces at arbitrary target electric potentials within the grand-canonical ensemble remain lacking.
	This gap means that electronic-level analysis of CP-MLMD trajectories still requires frame-by-frame DFT calculations, forming a bottleneck in the computational workflow.

	In this work, we propose a unified explicit electric potential graph neural network framework. A central design choice is that both models adopt a design philosophy in which the electric potential is treated as a scalar input feature and fused with atomic type embeddings at the first convolutional layer, allowing the model to incorporate potential information from the beginning. It uses high-accuracy, potential-labeled atomic configurations as training data and employs a decoupled dual-model architecture: PE-MACE, the MACE-based implementation of EEP-MLFF,\cite{Batatia2022a, Batatia2022Design} and PE-EDP for predicting real-space electron density distributions under arbitrary potentials, achieving a unified description from atomic-scale dynamics simulations to electronic-level density prediction.
	We take the Pt(111)/water interface as the benchmark system. 
	PE-MACE achieves high accuracy in energy and force predictions, and the RDFs from CP-MLMD simulations are in good agreement with CP-AIMD references. 
	Long-timescale CP-MLMD simulations at varying potentials also successfully reveal the potential-induced reorientation of interface water on Pt(111). 
	The PE-EDP model achieves an NMAE of 0.848\% on the test set, with planar-averaged electron densities and Bader charges in close agreement with DFT references.

\section{Methodology } 

\begin{figure}[H]
    \centering
    \includegraphics[width=6.5in]{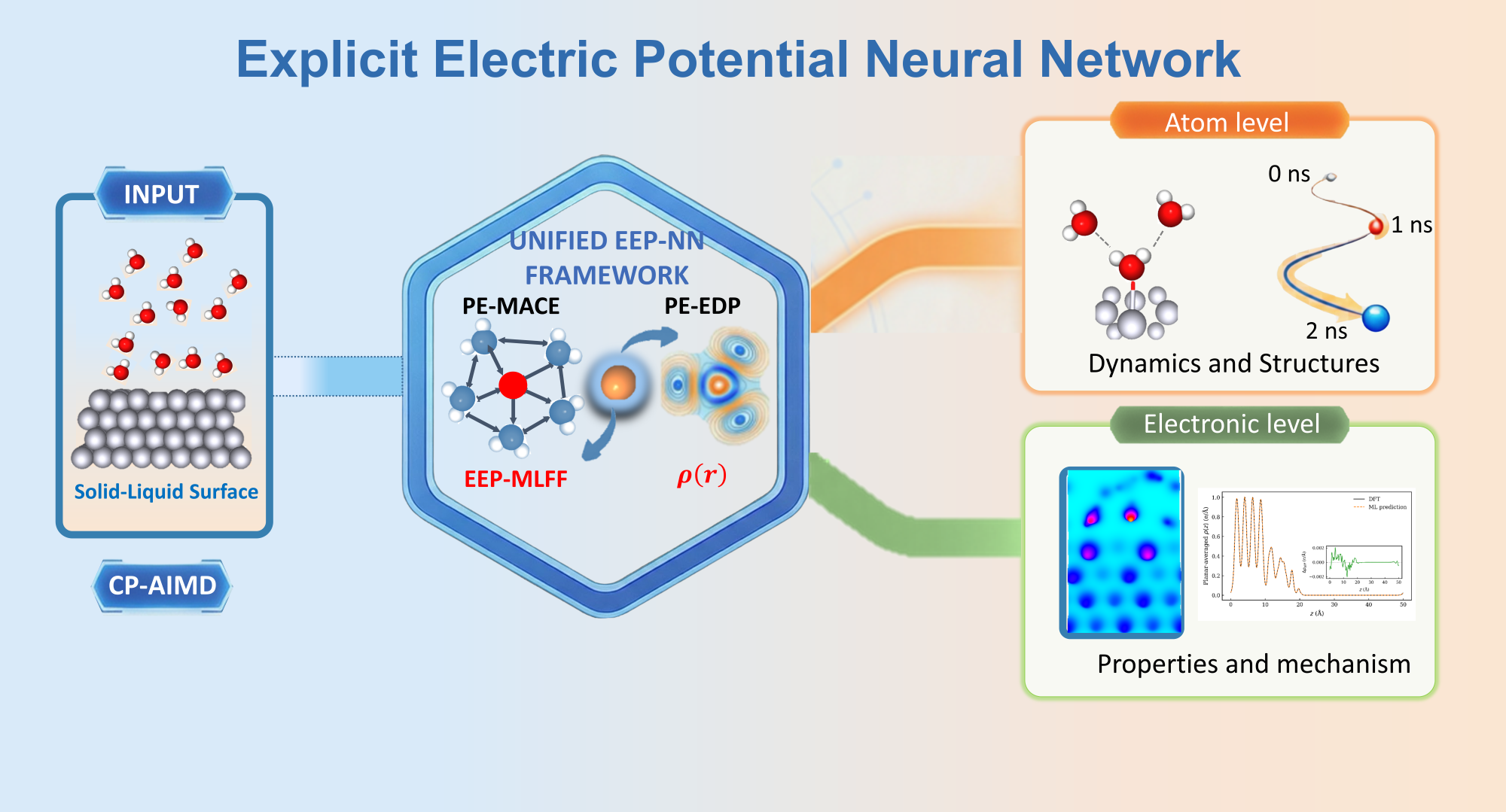}
    \caption{The overall architecture of EEP framework}
    \label{fig:EEP-NN_toc}
\end{figure}

The unified explicit electric potential framework consists of three components as shown in Figure~\ref{fig:EEP-NN_toc}. 
The first is a data generation module for producing DFT training datasets with embedded electric potential information. 
The second is a dual-model training module, wherein PE-MACE for atomic-scale dynamics and PE-EDP for electron density prediction are independently trained on the corresponding dataset. Both models are decoupled and incorporate a potential-embedding mechanism. 
The third is a model application stage that drives constant-potential molecular dynamics simulations and electron density predictions at arbitrary potentials, providing analytical tools for efficient simulation of electrochemical interfaces.
Both CP-AIMD and CP-MLMD in this work assume that electrons instantaneously relax to maintain the target potential at each ionic step. 
A discussion of this approximation and its comparison with potentiostat-type schemes is provided in the Supporting Information.

\subsection{Hy\_DFT software introduction}
Hy\_DFT (Grand Canonical DFT with Hybrid Solvation Model) is an interface package for performing constant-potential ab initio molecular dynamics (CP-AIMD) simulations. Built upon the Atomic Simulation Environment (ASE) framework,\cite{HjorthLarsen2017} it integrates VASP as the underlying quantum-chemical engine\cite{Kresse1993, Kresse1996} and employs VASPsol++ as the implicit electrolyte model to implement a hybrid explicit–implicit solvation scheme.\cite{Mathew2014, Mathew2019, Islam2023} 
VASPsol++ adopts a nonlocal cavity definition that effectively mitigates the solvent leakage artifact inherent in the original VASPsol,\cite{Islam2023} providing a more realistic physical description of electrochemical interfaces.

The constant-potential algorithm in Hy\_DFT follows the capacitance–Newton iteration framework proposed by Xia and Xiao.\cite{Xia2023} 
The central task is to establish a functional relationship between the number of electrons $N_e$ and the Fermi level $E_F$, and to iteratively update $N_e$ until $E_\mathrm{F}$ converges to a preset target value $E_{\mathrm{F}}^{\mathrm{target}}$, which relates to the electric potential. 
The work function is $\Phi = -E_{\mathrm{F}}$, was set to scale to the standard hydrogen electrode (SHE), for CP-AIMD, we follow the IUPAC-recommended absolute electrode potential of the standard hydrogen electrode ($U_{\mathrm{SHE}} = 4.44$ V),\cite{Trasatti+1986} the electric potential of the half-cell system is expressed as
\begin{equation}
    U (\mathrm{V vs. SHE}) = \Phi / e - 4.44 = -E_{\mathrm{F}} / e - 4.44.
    \label{eq:u_she}
\end{equation}
The interfacial differential capacitance $C_{\mathrm{diff}}$ relates changes in $N_e$ to changes in $E_{\mathrm{F}}$:
\begin{equation}
    C_{\mathrm{diff}} = \frac{\Delta N_e}{\Delta E_{\mathrm{F}}}.
    \label{eq:c_diff}
\end{equation}
Following the work of Xia and Xiao,\cite{Xia2023} the initial value of $C_{\mathrm{diff}}$ can be either supplied by the user or estimated from the electrode surface area:
\begin{equation}
    C_{\mathrm{diff}}^{(0)} = 
    \begin{cases}
        C_{\mathrm{diff}}^{\mathrm{initial}}, & \text{if provided}, \\
        A_{\mathrm{surface}} / 80, & \text{otherwise},
    \end{cases}
\end{equation}
where $A_{\mathrm{surface}} = \|\mathbf{a}_1 \times \mathbf{a}_2\|$ is calculated from the unit cell vectors.

To minimize the number of single-point calculations per ionic step, we adopt a three-stage iterative strategy that retains all computed data points to progressively improve the prediction accuracy. 
In the first iteration, $N_e$ is linearly extrapolated via the capacitance model:
\begin{equation}
    N_e^{(1)} = N_e^{(0)} + C_{\mathrm{diff}}^{(0)} \left( E_{\mathrm{F}}^{\mathrm{target}} - E_{\mathrm{F}}^{(0)} \right).
    \label{eq:stage1}
\end{equation}
When two data points become available, a linear fit of the $N_e$--$E_{\mathrm{F}}$ relationship is employed. 
Upon accumulating three or more data points, a quadratic polynomial $N_e(E_{\mathrm{F}}) = a_0 + a_1 E_{\mathrm{F}} + a_2 E_{\mathrm{F}}^2$ is fitted by least squares, and the target electron number is predicted as
\begin{equation}
    N_e^{(k+1)} = a_0 + a_1 E_{\mathrm{F}}^{\mathrm{target}} + a_2 \left( E_{\mathrm{F}}^{\mathrm{target}} \right)^2.
    \label{eq:stage3}
\end{equation}
In addition, a cross-step prediction strategy is introduced to further reduce the computational cost during CP-AIMD. 
The converged electron number, Fermi level, and updated differential capacitance from the previous ionic step are carried over to initialize the next step. 
The capacitance is updated as
\begin{equation}
    C_{\mathrm{new}} = \left| \frac{N_e^{(k)} - N_e^{(k-1)}}{E_{\mathrm{F}}^{(k)} - E_{\mathrm{F}}^{(k-1)}} \right|
    \label{eq:cap_update}
\end{equation}
and smoothed via an exponential moving average to suppress numerical oscillations:
\begin{equation}
    C_{\mathrm{diff}}^{(k+1)} = \alpha \cdot C_{\mathrm{new}} + (1-\alpha) \cdot C_{\mathrm{diff}}^{(k)},
    \label{eq:cap_smooth}
\end{equation}
where $\alpha$ is the smoothing factor. 
The initial electron number for the next ionic step is set to $N_e^{\mathrm{final}}$, and the subsequent prediction follows
\begin{equation}
    N_e^{\mathrm{next}} = N_e^{\mathrm{final}} + C_{\mathrm{diff}}^{\mathrm{final}} \left( E_{\mathrm{F}}^{\mathrm{target}} - E_{\mathrm{F}}^{\mathrm{final}} \right).
    \label{eq:cross_step}
\end{equation}

This strategy can potentially allow the next ionic step to converge to the target Fermi level in one or two SCF cycles, thereby significantly lowering the overall computational cost.

Beyond CP-AIMD, Hy\_DFT also supports constant-potential single-point optimization, constant-charge AIMD, enhanced-sampling simulation using Plumed, and PE-MACE-driven CP-MLMD simulations within ASE. The package adopts a modular architecture managed through YAML configuration files, allowing flexible combination of independent functional modules. 
 
Importantly, Hy\_DFT serves a dual role as both a simulation engine and a training-data generator for subsequent machine learning models. 
During CP-AIMD execution, the converged electric potential $U$ at each ionic step is automatically embedded into the atomic structure as a \texttt{field\_scalar} attribute and exported in the Extended XYZ format. 
The resulting atomic data $(\mathbf{R}, U)$ can be directly used to train the potential-embedded machine learning models (PE-MACE and PE-EDP) described below, without additional data post-processing. 
The source code of Hy\_DFT is publicly available at \url{https://github.com/xke123502/Hybrid_GCDFT}.

\subsection{PE-MACE model architecture}

\begin{figure}[H]
    \centering
    \includegraphics[width=6.5in]{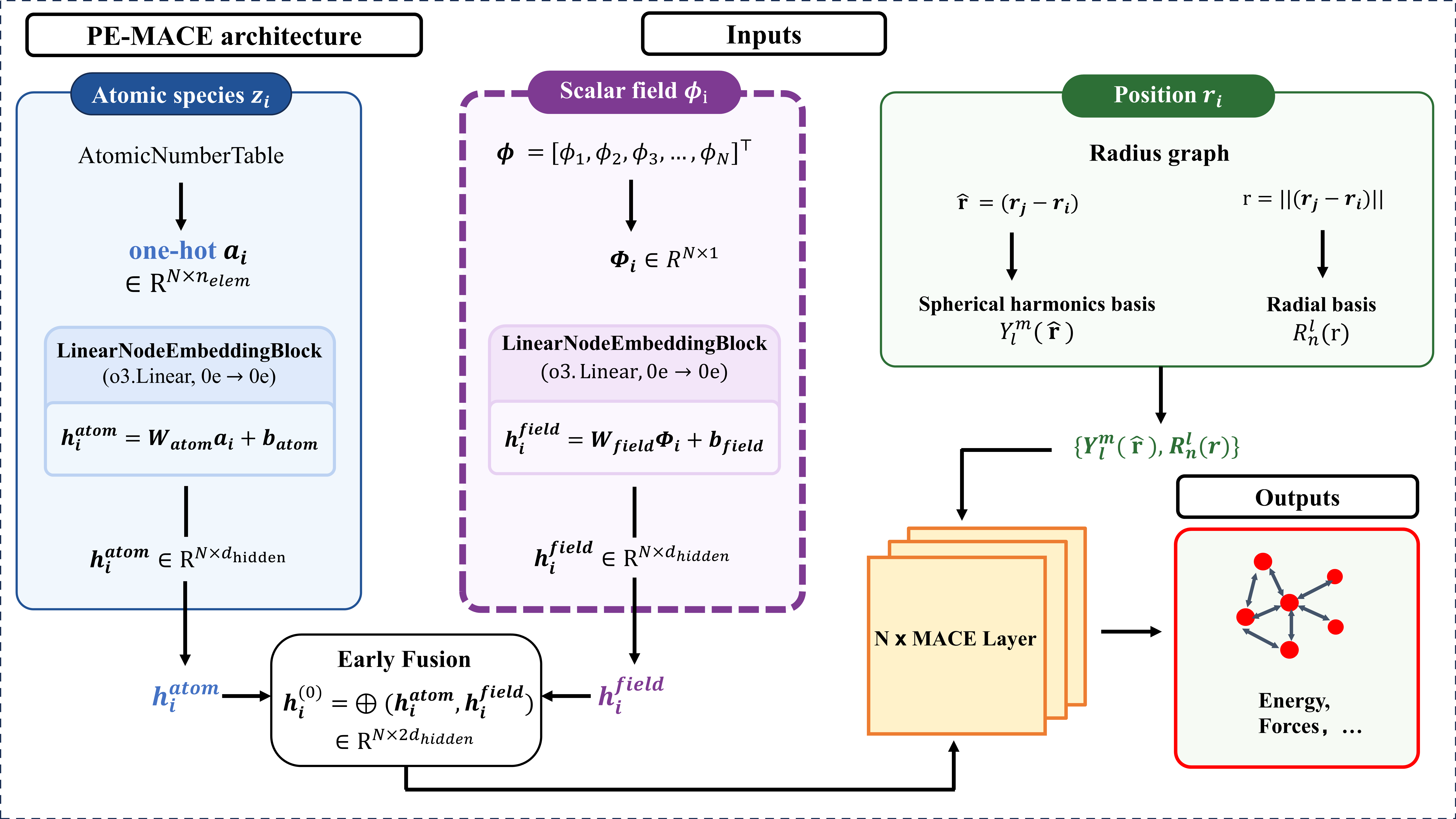}
    \caption{Schematic of the PE-MACE model architecture.}
    \label{mace}
\end{figure}

PE\_MACE (Potential-Embedded MACE) is an implementation of the explicit electric potential machine learning force field (EEP-MLFF) based on the MACE framework.\cite{Batatia2022Design, Batatia2022a}
The model builds upon the E(3)-equivariant neural network architecture of MACE and introduces a scalar field embedding mechanism to explicitly incorporate the external electric potential, thereby describing the influence of the electric potential on interatomic interactions in electrochemical systems.

As shown in Figure~\ref{mace}, PE\_MACE takes as input the atomic positions $\mathbf{R} = \{\mathbf{r}_i\}_{i=1}^N$, atomic species $\mathbf{Z} = \{z_i\}_{i=1}^N$, and a scalar field $\phi = \{\phi_i\}_{i=1}^N$, where $\phi_i$ is the electric potential of atom $i$, and $\phi$ takes the same value for all atoms, corresponding to the electric potential of the system.
The atomic positions are used to construct the edge index set $\mathcal{E} = \{(i,j)\}$, which defines atomic pairs within a cutoff distance $r_{ij} < r_{\max}$.

The central modification with respect to the original MACE lies in the early fusion of node features.
For each atom $i$, the atomic species $z_i$ is mapped to an initial node feature vector $\mathbf{h}_i^{\text{atom}} = \mathrm{LinearEmbedding}(z_i) \in \mathbb{R}^{d_{\text{hidden}}}$ via a linear embedding layer, where $d_{\text{hidden}}$ is the hidden feature dimension.
Simultaneously, the scalar field value $\phi_i$ is mapped through a separate linear embedding layer to a feature vector of the same dimension, $\mathbf{h}_i^{\text{field}} = \mathrm{LinearEmbedding}(\phi_i) \in \mathbb{R}^{d_{\text{hidden}}}$.
The two embeddings are concatenated along the feature dimension to form the augmented initial node feature:
\begin{equation}
    \mathbf{h}_i^{(0)} = [\mathbf{h}_i^{\text{atom}},\; \mathbf{h}_i^{\text{field}}] \in \mathbb{R}^{2d_{\text{hidden}}}.
    \label{eq:node_feature}
\end{equation}
Projecting the scalar potential to the same dimension as the chemical embedding before concatenation ensures that the potential information carries comparable weight in the resulting feature representation, preventing it from being overwhelmed by the higher-dimensional chemical features.
This early fusion strategy allows the model to be aware of the electric potential from the very first interaction layer, facilitating the learning of potential-dependent interatomic interactions.
 
The electric potential is treated as a scalar field input and accordingly adopts the $0e$ irreducible representation (i.e., a scalar under rotations), which preserves the rotational equivariance of the model, since a scalar is invariant under rotations.
This design enables the model to describe the influence of the electric potential on the system energy while maintaining physical consistency.
 
Edge features are constructed by encoding both distance and directional information.
The interatomic distance is encoded through a radial embedding module that combines a Bessel-type radial basis with a smooth polynomial cutoff function:\cite{Batatia2022Design}
\begin{equation}
    \mathbf{f}^{\text{radial}}_{ij} = \sqrt{\frac{2}{r_{\max}}}
    \frac{\sin\!\left(\dfrac{n\pi\, r_{ij}}{r_{\max}}\right)}{r_{ij}}
    \cdot f_{\text{cutoff}}(r_{ij}),
    \label{eq:radial_embedding}
\end{equation}
where $n = 1, 2, \ldots, N_{\text{radial}}$, $f_{\text{cutoff}}(r_{ij})$ is the smooth cutoff function, and $N_{\text{radial}}$ is the number of radial basis functions.
The directional information of the edge vector is encoded through spherical harmonics to maintain rotational equivariance:
\begin{equation}
    \mathbf{Y}_{ij} = \{Y_{\ell m}(\hat{\mathbf{r}}_{ij})\}_{\ell=0,\, m=-\ell,\, \ldots,\, \ell}^{\ell_{\max}},
    \label{eq:spherical_harmonics}
\end{equation}
where $\hat{\mathbf{r}}_{ij} = \mathbf{r}_{ij}/r_{ij}$ is the normalized direction vector and $\ell_{\max}$ is the maximum angular momentum quantum number.
 
The above structural and potential features serve as the initial input to $L$ MACE interaction layers.\cite{Batatia2022a, Batatia2022Design, Kovacs2023}
Each interaction layer consists of an interaction block followed by an equivariant product block.
In the interaction block, for each edge $(i,j)$, the neighbor node features $\mathbf{h}_j^{(l)}$ and the spherical harmonics $\mathbf{Y}_{ij}$ are combined via an equivariant tensor product, weighted by the output of a radial MLP, and aggregated over all neighbors.
A self-connection simultaneously generates a residual feature:
\begin{equation}
    \mathbf{m}_i^{(l)},\; \mathbf{sc}_i^{(l)} = \mathrm{InteractionBlock}\!\left(\mathbf{h}_i^{(l)},\, z_i,\, \{\mathbf{Y}_{ij},\, \mathbf{f}_{ij}^{\text{radial}}\}_{j \in \mathcal{N}(i)}\right).
    \label{eq:message_aggregation}
\end{equation}
The aggregated message and the residual feature are then fused through an equivariant product block to update the node features:
\begin{equation}
    \mathbf{h}_i^{(l+1)} = \mathrm{EquivariantProduct}\!\left(\mathbf{m}_i^{(l)},\, \mathbf{sc}_i^{(l)},\, z_i\right),
    \label{eq:node_update}
\end{equation}
where the equivariant product block employs a symmetric contraction to construct many-body correlation features, combined with the residual term $\mathbf{sc}_i^{(l)}$, ensuring that the updated node features maintain E(3) equivariance.
 
The site energy of each atom is obtained as a sum of readout contributions from all layers,
\begin{equation}
    E_i = \sum_{l=1}^{L} \mathcal{R}^{(l)}\!\left(\mathbf{h}_i^{(l)}\right),
    \label{eq:site_energy}
\end{equation}
where the readout function $\mathcal{R}^{(l)}$ acts on the rotationally invariant ($l=0$) components of the node features. $\mathcal{R}^{(l)}$ is a linear map for all layers except the last, for which a shallow MLP is used to preserve the body-ordered nature of the energy decomposition.\cite{Batatia2022a}
The total potential energy is $E_{\text{total}} = \sum_i E_i$, and atomic forces are obtained by automatic differentiation,
\begin{equation}
    \mathbf{F}_i = -\frac{\partial E_{\text{total}}}{\partial \mathbf{r}_i}.
    \label{eq:force}
\end{equation}
 
PE\_MACE preserves the E(3) equivariance of the original MACE architecture, ensuring that model outputs (energies, forces, and stresses) transform correctly under three-dimensional Euclidean transformations (rotations and translations).
By taking the Fermi level or the equivalent electric potential as an input feature, the model learns the potential energy surface under varying potential conditions, enabling long-time-scale simulations of electrochemical interface processes.
The CP-MLMD simulations are carried out using a modified ML-MACE plugin in LAMMPS,\cite{Thompson2022} which has been released together with the PE-MACE source code at \url{https://github.com/xke123502/PE_MACE}.

\subsection{PE-EDP model architecture}
\begin{figure}[H]
	\centering
	\includegraphics[width=6.5in]{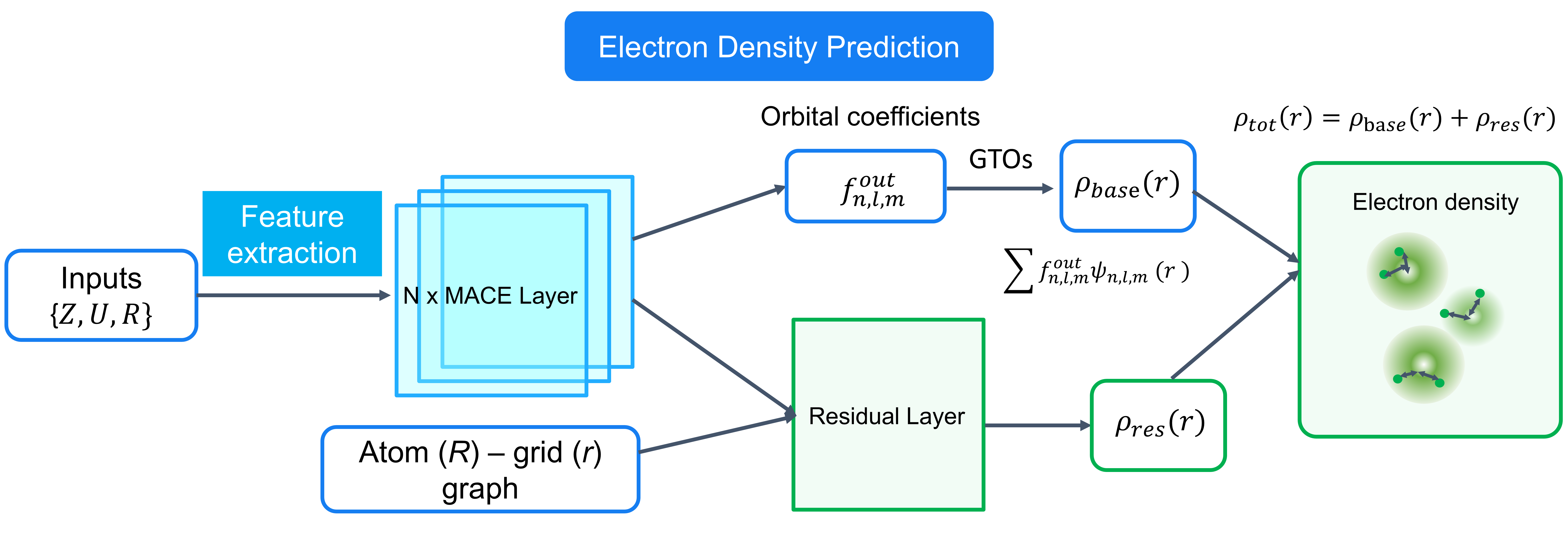}
	\caption{Schematic of the PE-EDP model architecture.}
	\label{PE-EDP}
\end{figure}
The electron density $\rho(\mathbf{r})$ is the central quantity in density functional theory (DFT), from which all physicochemical properties of a system can, in principle, be determined.\cite{Kohn1965, RevModPhys.71.1253}
However, conventional Kohn--Sham DFT calculations scale as $O(N^3)$, which severely limits their applicability to large-scale systems and long-time simulations.
To address this accuracy--efficiency trade-off, various neural network architectures have been developed to map directly from atomic coordinates to the ground-state electron density, thereby bypassing the expensive self-consistent field (SCF) iterations. Prior research has mainly focused on two approaches to predicting the electron density. 
Grid-based methods treat all real-space grid points as probe nodes and predict the charge density through message passing between atoms and these probe nodes.
In models such as DeepDFT and ChargE3Net, \cite{jorgensen_equivariant_2022-1, Koker2024b}
the system is formulated as an atom-probe graph, where atoms and grid points serve as two types of nodes, the model first learns the chemical environment via atom--atom message passing, then predicts the density at each probe point via atom-to-probe message passing.
Although grid-based methods are flexible and can achieve high accuracy, they incur substantial computational costs, even for small systems the number of grid points is typically very large, and the atom--probe connectivity leads to rapidly increasing computational demands upon multi-layer message passing. Another class of methods is orbital-based approaches, which do not directly predict the density value at grid points but instead represent it as a linear combination of the basis functions.\cite{Cheng2023infgcn, kim2024gaussian, fu2024recipe}
Only the expansion coefficients for each atom need to be predicted.
The full three-dimensional density is then reconstructed by superposing all basis functions and evaluating them on the spatial grid.
This formulation scales favorably to large systems and offers substantially faster inference than grid-based methods.

As shown in Figure~\ref{PE-EDP}, PE-EDP (Potential-Embedded Electron Density Prediction) is an equivariant graph neural network framework for predicting real-space electron density distributions under arbitrary electric potential conditions.
The density prediction follows an orbital-based scheme: multiple MACE interaction layers perform equivariant message passing to learn the local atomic environment and predict the basis set expansion coefficients for each atom.
In practice, only a finite number of basis functions can be used, and the radial part of the GTO basis is neither complete nor orthogonal, which limits the expressiveness of the coefficient-based density representation. \cite{Cheng2023infgcn}
To mitigate this issue, the equivariant residual layer operating on the grid is introduced as a residual correction to the predicted density.
Following the same strategy as in PE-MACE, the model incorporates the electric potential ($U$) through an early fusion embedding of atomic species and scalar field features.
The model takes as input the atomic species, atomic coordinates, the electric potential (or equivalently, the Fermi level) as a scalar field, and a set of sampling grid points $\mathcal{G} = \{\mathbf{g}_k\}$.
The initial node features are formed by concatenating the atomic species embedding and the scalar field embedding, $\mathbf{h}_i^{(0)} = [\mathbf{h}_i^{\text{atom}},\, \mathbf{h}_i^{\text{field}}]$.
An atomic neighborhood graph is constructed within a cutoff radius $r_c$, with the edge set $\mathcal{E} = \{(i,j) : \|\mathbf{r}_j - \mathbf{r}_i\| < r_c\}$.
The interatomic distances are expanded in a Gaussian radial basis, and the edge directions are encoded by spherical harmonics, the features are then processed through MACE interaction layers for tensor product operations and message aggregation.
These steps are identical to those in PE-MACE.
 
The model employs Gaussian-type orbital (GTO) basis functions\cite{Cheng2023infgcn,fu2024recipe}
\begin{equation}
    \psi_{n\ell m}(\mathbf{r}) = \mathcal{N}_{n\ell}\, \exp(-\alpha_n r^2)\, r^\ell\, Y_\ell^m(\hat{\mathbf{r}}),
    \label{eq:gto}
\end{equation}
where the angular momentum quantum number $\ell$ takes non-negative integer values and the magnetic quantum number $m$ ranges from $-\ell$ to $\ell$, giving $2\ell+1$ components for each $\ell$.
$\mathcal{N}_{n\ell}$ is a normalization constant.
For the Gaussian exponents $\alpha_n$, the model adopts a logarithmic discretization scheme rather than a conventional linear spacing.
This provides an adaptive radial resolution, denser basis functions are allocated in the short-range region where interatomic interactions vary rapidly, while sparser sampling is used in the long-range region where interactions decay smoothly, matching the physical characteristics of interatomic interactions.
 
The node features $\mathbf{h}_i^{(L)}$ learned by the MACE layers serve as the expansion coefficients for each basis function, while the orbital functions $\psi_{n\ell m}$ provide the spatial distribution.
The density contribution from atom $i$ to grid point $k$ is obtained by the element-wise weighted summation of the two, and the contributions from all atoms are aggregated at each grid point to yield the orbital-based density.
Adding the residual correction gives the final electron density:
\begin{equation}
    \rho(\mathbf{g}_k) = \sum_{i=1}^{N} \sum_{n\ell m} \psi_{n\ell m}(\mathbf{g}_k - \mathbf{r}_i)\, [\mathbf{h}_i^{(L)}]_{n\ell m} + \rho^{\text{res}}(\mathbf{g}_k).
    \label{eq:density}
\end{equation}
 
PE-EDP employs E(3)-equivariant representations and tensor products in both the atomic graph message passing and the grid-based residual mapping.
Consequently, the learned node features and the scalar residual both transform correctly under rotations and translations.
The orbital basis $\psi_{n\ell m}$ satisfies rotational equivariance through its spherical harmonic component, while the radial part is rotationally invariant.
The density $\rho(\mathbf{g}_k)$ is a scalar field and transforms as such under global rotations and translations.
The Fermi-level embedding adopts the $0e$ scalar irreducible representation and therefore does not break the equivariance of the model.
 
To reduce GPU memory usage during training, we adopt a sparse sampling strategy in which only a small number of grid points are randomly sampled at each training step.
This approach reduces the memory footprint while maintaining the generalization capability of the model.

\section{Results and discussions}
The structure and dynamics of interface water molecules determine mass transport, charge distribution, and reaction processes.
Prior studies have shown that the hydrogen-bond network and spatial orientation of interface water can differ substantially from those in bulk.\cite{Wang2024}
Such local water structures may significantly regulate ion adsorption, charge transfer, and reaction pathways.\cite{Rizo2015,Zhong2022} 
However, it remains unclear which water structures dominate above processes under what conditions for specific solid–liquid interfaces.\cite{Wang2024}
In this work, we use the Pt(111)/water interface as a representative system to present the training and validation of an EEP-MLFF model and a PE-EDP model, and to provide a preliminary characterization of potential-induced changes in interface water structure and dynamics. 
At present, we consider only electric potentials at or below the potential of zero charge (PZC), a more detailed analysis will be reported in future work.

\subsection{EEP-MLFF for the Pt(111)/Water Interface}
\begin{figure}[H]
	\includegraphics[width=6.5in]{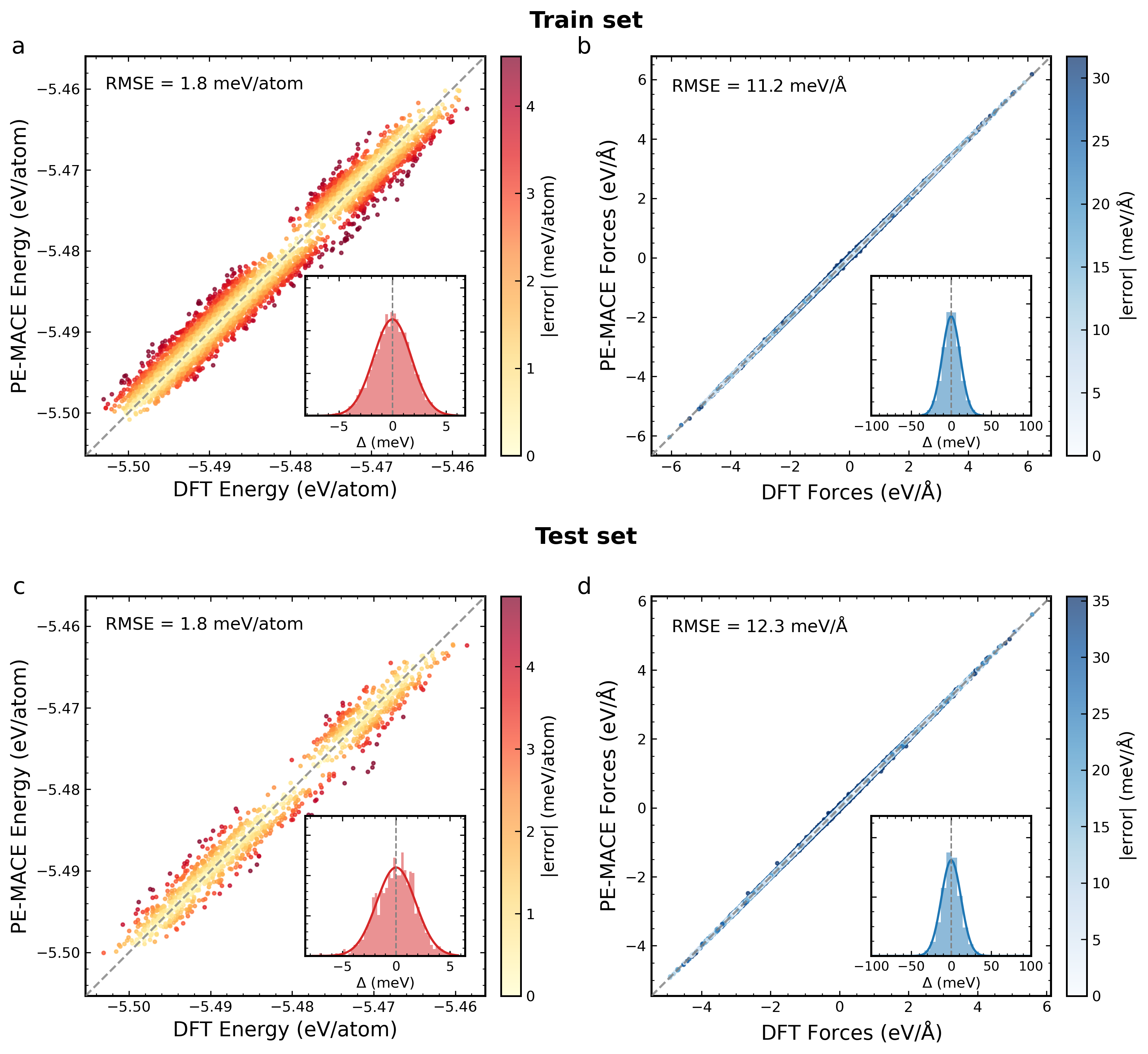}
	\caption{Comparison of the PE-MACE model accuracy on the training and test sets. (a, b) Training set; (c, d) test set. The left panels show energies (per atom), and the right panels show atomic forces. 
	Point color encodes the absolute error $|y-x|$ (meV or meV/\AA).
	Insets display the distribution of residuals $\Delta$ (in meV), defined as the difference between PE-MACE predictions and DFT references}
\label{fig1}
\end{figure}

We develop the potential-embedded MACE (PE-MACE) model to train EEP-MLFF for the Pt(111)/water interface, based on the open-source machine learning force field software MACE.\cite{Batatia2022a, Batatia2022Design}
The dataset is generated by Hy\_DFT from CP-AIMD trajectories of $10^4$ steps at each of three electric potentials ($-0.44$, $-0.04$, and $+0.26$ V), and is partitioned into training, validation, and test sets of 4562, 240, and 1201 structures, respectively.
The model employs 3 MACE interaction layers with a maximum correlation order of 2. more detailed training hyperparameters are listed in Table~S1.
Table~S2 reports the root-mean-square errors on the training, validation, and test sets. As shown in Figure~\ref{fig1}, the model achieves high accuracy in both energy and force predictions. On the test set, the RMSE values of energy and forces are 1.8 meV/atom and 12.3 meV/\AA, respectively, with a relative force RMSE of 1.93\%, confirming that the model reproduces the potential energy surface with satisfactory accuracy and is sufficiently accurate for subsequent CP-MLMD simulations.

To further validate the accuracy of PE-MACE, we compare 10 ps trajectories from CP-AIMD and CP-MLMD simulations starting from the same initial configuration. Although this timescale is too short for full equilibration, it is sufficient to assess the consistency of short-time dynamical behavior between the two approaches. It can be seen from Figures~S1--S3 that the O--O, O--H, and H--H radial distribution functions at all three potentials ($-0.44$, $-0.04$, and $+0.26$ V) are in good agreement with the CP-AIMD references, confirming the accuracy of PE-MACE in dynamical simulations.

Since the present CP-AIMD trajectories are too short to characterize potential-induced reorientation of interface water with statistical reliability, we perform 4 ns CP-MLMD simulations driven by PE-MACE to obtain statistically converged orientation distributions.
Before orientation analysis, we estimate uncertainty in the trained PE-MACE model with a committee of three models trained independently (different random seeds).\cite{Schran2020, Zhang2019c} The maximum inter-model variance of atomic forces along CP-MLMD trajectories at each potential (every 100 steps) is ~0.03 eV/\AA\ (Figure~S4), well below the threshold typically used to trigger active-learning sampling,\cite{Gong2024} indicating adequate coverage of equilibrium configurations in the electric potential window studied here without an additional active-learning iteration.

\begin{figure}[H]
	\centering
	\includegraphics[width=6.5in]{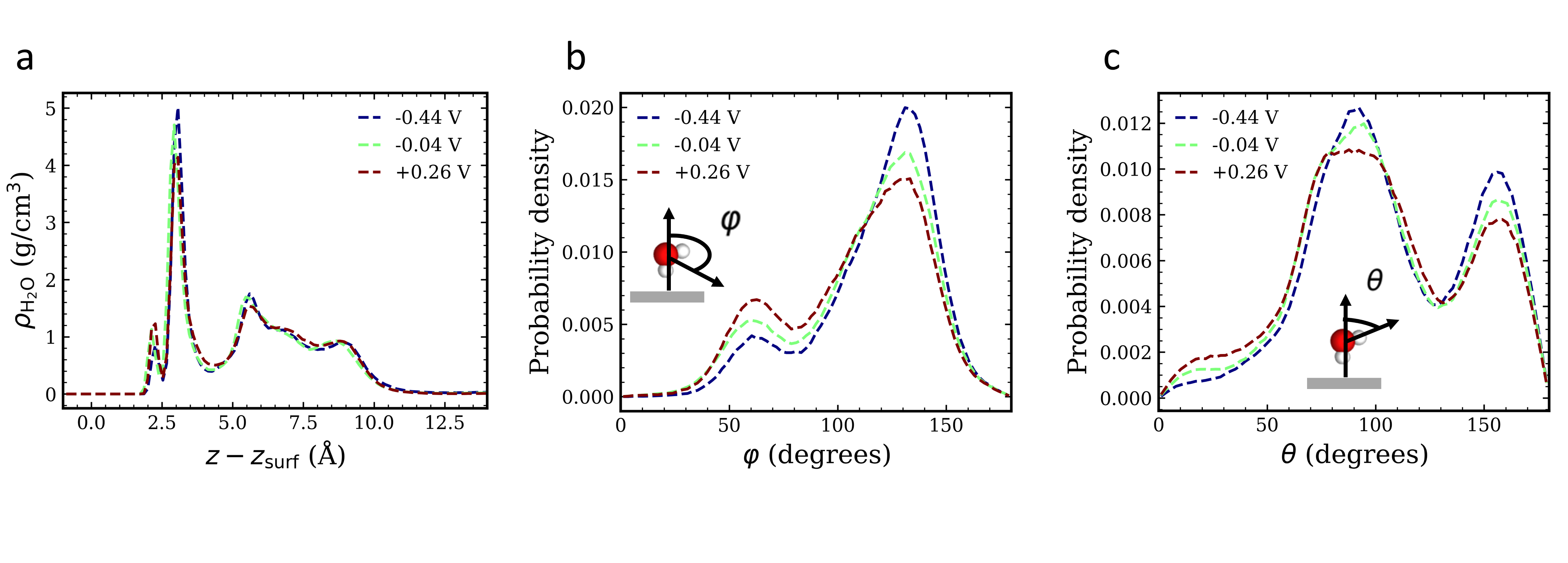}
	\caption{Potential-dependent structure of interfacial water at the Pt(111)/water interface from 4~ns CP-MLMD simulations.
	(a) Water density profile along the surface normal.
	(b) Probability density distribution of the bisector angle $\varphi$, defined as the angle between the $\mathrm{O}\!\rightarrow\!\mathrm{H\!H}$ midpoint direction and the surface normal.
	(c) Probability density distribution of the $\mathrm{O}$--$\mathrm{H}$ bond angle $\theta$ with respect to the surface normal.
	The orientation analyses in (b) and (c) include only first-layer water molecules, identified by the first minimum of the density profile in (a).
	All potentials are referenced to the standard hydrogen electrode (SHE).}
	\label{fig:orientation}
\end{figure}

Next, we analyze the orientational structure of interface water. To quantify potential-induced reorientation, we average over the last 3.5 ns of each equilibrated trajectory.
Figure~\ref{fig:orientation}(a) shows the oxygen density profiles of interfacial water along the surface normal at three electric potentials. 
All three curves are largely insensitive to the applied potential, however, systematic and subtle variations can be directly observed, as the potential becomes more negative, the first-layer oxygen density peak is slightly reduced while the second-layer peak is slightly enhanced, a trend geometrically consistent with the potential-induced strengthening of the H-down orientation (Figure~\ref{fig:orientation}(b, c)). 
The distribution of the bisector angle $\varphi$ (Figure~\ref{fig:orientation}b) features a dominant peak at $\varphi \approx 130^\circ$ and a weaker shoulder near $\varphi \approx 70^\circ$.  
The peak at $\varphi \approx 130^\circ$, corresponding to the H-down orientation in which the molecular bisector tilts toward the electrode surface, increases monotonically in intensity as the potential becomes more negative, while the contribution near $\varphi \approx 70^\circ$ (H-up orientation) is correspondingly reduced. 
Figure~\ref{fig:orientation}(c) shows the O--H bond-angle distribution for $\theta$.
A primary peak near $90^\circ$ persists at all potentials, indicating that one O--H bond remains approximately parallel to the surface to sustain lateral hydrogen bonding.
A secondary peak near $160^\circ$, corresponding to the other O--H bond pointing toward the electrode, intensifies with decreasing potential.

The combined $\varphi$ and $\theta$ distributions reveal a consistent molecular picture: one O--H bond lies parallel to the surface, participating in the hydrogen-bond network, while the other O--H bond undergoes a potential-dependent change in orientation.
This trend is consistent with the potential-dependent orientational behavior reported in the literature~\cite{Le2020a, Garcia-Araez2009}. At more negative potentials, the higher surface electron density electrostatically attracts the partially positive H atoms toward the electrode, favoring H-down ordering, while increasing the potential toward the PZC progressively weakens this preference. The systematic potential-dependence of the orientational distributions demonstrates that the PE-MACE force field can reproduce the potential-induced reorientation of interfacial water molecules.

\subsection{Potential-embedded Electron Density Prediction for the Pt(111)/Water Interface}
We constructed the PE-EDP model to predict real-space electron densities under arbitrary electric potential conditions for the Pt(111)/water interface.
The training data were extracted from the CP-AIMD trajectories at three electric potentials ($-0.44$, $-0.04$, and $+0.26$ V vs.\ SHE), with 1000 structures per potential, yielding a total of 3000 charge density samples.
The training, validation, and test sets contain 2250, 450, and 300 data points, respectively.
The model hyperparameters are summarized in Table~S3.

Following the metric adopted in previous work,\cite{fu2024recipe, jorgensen_equivariant_2022-1, Cheng2023infgcn} we evaluate the prediction accuracy using the normalized mean absolute error (NMAE).
\begin{equation}
	\text{NMAE}(\hat{\rho}) = \frac{\int_{\mathbb{R}^3} |\rho(\mathbf{r}) - \hat{\rho}(\mathbf{r})| \, \mathrm{d}V}{\int_{\mathbb{R}^3} |\rho(\mathbf{r})| \, \mathrm{d}V}
	\end{equation}
The NMAE values for the training, validation, and test sets are 0.846\%, 0.851\%, and 0.848\%, respectively, all below 0.9\%.
The nearly identical errors across the three splits indicate that the model does not overfit.
It is worth noting that, to alleviate GPU memory constraints, only 1024 grid points (approximately 0.02\% of the total) were randomly sampled per training step, yet the model still reconstructs the full three-dimensional electron density distribution with high fidelity.

\begin{figure}[H]
    \centering
    \includegraphics[width=3.5in]{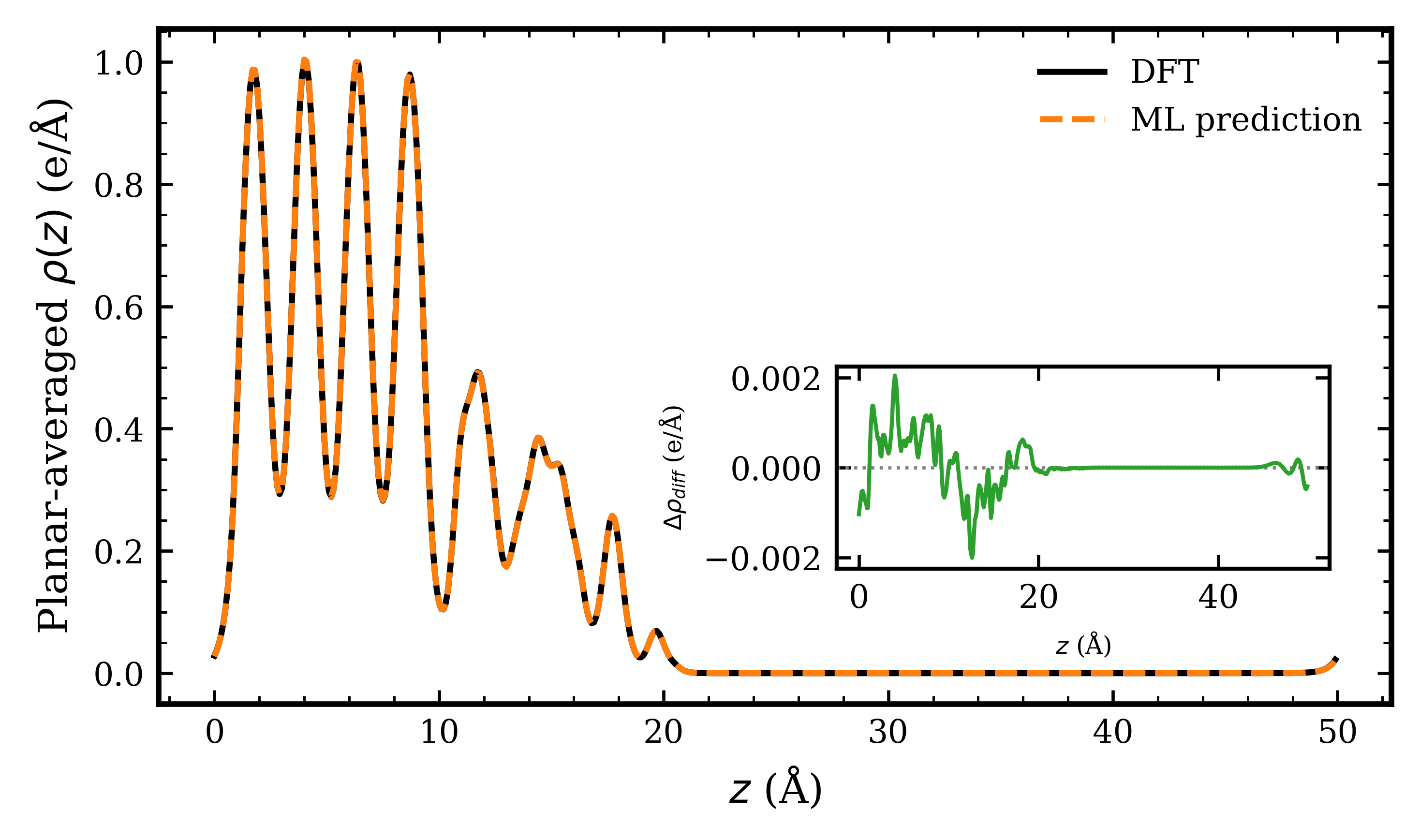}
    \caption{Comparison of the planar-averaged electron density $\bar{\rho}(z)$ along the surface normal direction between DFT (black solid line) and PE-EDP prediction (orange dashed line) for a representative Pt(111)/water configuration. The inset shows the difference $\Delta\bar{\rho}(z) = \bar{\rho}_{\text{ML}}(z) - \bar{\rho}_{\text{DFT}}(z)$.}
    \label{fig:planar_averaged_electron_density}
\end{figure}

To further assess the quality of the predicted electron density, we randomly selected a Pt(111)/water configuration and compared the DFT and PE-EDP results in detail.
Figure~\ref{fig:planar_averaged_electron_density} shows the planar-averaged electron density $\bar{\rho}(z)$ along the surface normal direction.
The ML prediction agrees closely with the DFT reference across the entire simulation cell.
In the $z = 0$--10~\AA{} region, four sharp peaks corresponding to the four Pt layers are well reproduced in both peak position and shape.
In the $z = 10$--20~\AA{} region of the explicit water layer, the predicted density also reflects the spatial distribution of O and H atoms, with peak heights decreasing with increasing distance from the surface.
The inset shows the difference $\Delta\bar{\rho}(z) = \bar{\rho}_{\text{ML}}(z) - \bar{\rho}_{\text{DFT}}(z)$, which remains within $\pm$0.002~e/\AA{} throughout the cell.
 
Bader charge analysis on the same structure (Figure~S5) shows that the ML-predicted atomic charges agree closely with the DFT values.
Among the three elements, H atoms exhibit the smallest prediction errors, followed by O atoms, while Pt atoms show relatively larger deviations.

Since public electron density datasets with explicit electric potential information are scarce, we further evaluated our electron density prediction model without the potential embedding module on three public benchmarks (QM9, MD, and Cubic) to assess the general capability of the underlying architecture. Dataset definitions and splits are described in the Supporting Information.

\begin{table}[H]
	\centering
	\caption{Comparison of our electron density prediction model with other models on public datasets (NMAE \%)}
	\label{tab:public_dataset_comparison}
	\begin{tabular}{lccccc}
		\hline
		Dataset & Our model & ELECTRA \cite{Elsborg2026} & SCDP \cite{fu2024recipe} & GPWNO \cite{kim2024gaussian} & InfGCN \cite{Cheng2023infgcn} \\
		\hline
		MD-ethanol       & 1.87 & \textbf{1.02} & $2.34\pm0.25$ & 4.00 & 8.43  \\
		MD-benzene       & 1.02 & \textbf{0.45} & $1.13\pm0.06$ & 2.45 & 5.11  \\
		MD-phenol        & 1.41 & \textbf{0.56} & $1.29\pm0.07$ & 2.68 & 5.51  \\
		MD-resorcinol    & 1.44 & \textbf{0.62} & $1.35\pm0.08$ & 2.73 & 5.95  \\
		MD-ethane        & 1.40 & \textbf{0.91} & $2.05\pm0.12$ & 3.67 & 7.01  \\
		MD-malonaldehyde & 2.42 & \textbf{0.80} & $2.71\pm0.60$ & 5.32 & 10.34 \\
		QM9              & 0.31 & \textbf{0.176} & $0.178\pm0.002$ & 0.73 & 0.93  \\
		Cubic            & 2.72 & --            & \textbf{$2.59\pm0.25$}   & 7.69 & 8.98  \\
		\hline
	\end{tabular}
  \end{table}

As shown in Table~\ref{tab:public_dataset_comparison}, the model achieves accuracy comparable to the SCDP model\cite{fu2024recipe} on the Cubic dataset, which, like our Pt/water system, consists of periodic structures.
On the QM9 and MD molecular datasets, a gap remains relative to the ELECTRA model.\cite{Elsborg2026}
This is likely because our current model fixes the orbital centers at atomic positions, whereas ELECTRA employs floating orbitals that provide a more flexible density representation without increasing the basis set size.
Future work will explore architectural improvements to narrow this gap.

\section{Conclusion} 

In conclusion, we have presented a unified potential-embedded machine learning framework that combines three components for simulating electrochemical interfaces at both the atomic and electronic levels.
Hy\_DFT performs constant-potential ab initio molecular dynamics and automatically generates training data with the electric potential embedded as a scalar field attribute.
PE-MACE and PE-EDP, two decoupled equivariant graph neural network models sharing the same early fusion potential embedding strategy, predict atomic forces and real-space electron densities, respectively, under arbitrary electric potential conditions.
 
Using the Pt(111)/water interface as a test system, the PE-MACE model achieves an energy RMSE of 1.8 meV/atom and a force RMSE of 12.3 meV/\AA on the test set.
Radial distribution functions from CP-MLMD trajectories agree well with CP-AIMD references at three electric potentials.
Long-time (4~ns) CP-MLMD simulations reveal the potential-induced reorientation of interface water molecules, with the dipole orientation distribution shifting systematically toward the H-down configuration at negative electric potentials, consistent with previously reported trends.
The PE-EDP model achieves an NMAE of 0.848\% on the test set, and the predicted planar-averaged electron density and Bader charges are in close agreement with DFT references, providing quantitative electronic-level information that complements the atomic-scale description of PE-MACE.

\begin{suppinfo}
The Supporting Information is available free of charge.
\begin{itemize}
	\item SI.pdf: 
		Computation details of DFT calculations, the structure and parameters of the neural network for the machine-learning model, as well as the parameters for implementing MD simulations.
		Supplementary Figures and Tables.
	\end{itemize}

\end{suppinfo}

\begin{acknowledgement}
This work was supported by the National Key Research and Development Programs of China (2022YFA1503103, 2023YFA1506902) and the National Natural Science Foundation of China NSF (Grand No. 22073041). We thank the High Performance Computing Center of Nanjing University for computational resources. 
	
\end{acknowledgement}

\bibliography{ref}
	
\end{document}